\documentclass[conference]{IEEEtran}
\IEEEoverridecommandlockouts
\usepackage{cite}
\usepackage{amsmath,amssymb,amsfonts}
\usepackage{algorithmic}
\usepackage{graphicx}
\usepackage{textcomp}
\usepackage{xcolor}
\usepackage{orcidlink}
\usepackage{stfloats}
\usepackage{siunitx}
\usepackage{array}
\usepackage{tabularray}
\usepackage{multirow}
\def\BibTeX{{\rm B\kern-.05em{\sc i\kern-.025em b}\kern-.08em
    T\kern-.1667em\lower.7ex\hbox{E}\kern-.125emX}}
\begin{document}

\title{A Data-Driven Condition Monitoring Method for Capacitor in Modular Multilevel Converter (MMC)
}

\author{
Shuyu Ou$^\ast$\orcidlink{0000-0002-6339-6984},
Mahyar Hassanifar$^\dagger$\orcidlink{0000-0003-1506-6312},
Martin Votava$^\ddagger$\orcidlink{0000-0001-5762-7646},
Marius Langwasser$^\dagger$\orcidlink{0000-0001-9909-7516}, 
Marco Liserre$^\dagger$$^\ddagger$\orcidlink{0000-0002-0818-2684},  \\
Ariya Sangwongwanich$^\ast$\orcidlink{0000-0002-2587-0024},
Subham Sahoo$^\ast$\orcidlink{0000-0002-7916-028X}, and
Frede Blaabjerg$^\ast$\orcidlink{0000-0001-8311-7412} \\

$^\ast$ Department of Energy, 
 Aalborg University, Aalborg, Denmark\\
 $^\dagger$ Chair of Power Electronics, 
 Kiel University, Kiel, Germany\\
 $^\ddagger$ Electronic Energy Systems, 
 Fraunhofer Institute for Silicon Technology ISIT, Kiel, Germany\\
so@energy.aau.dk, mha@tf.uni-kiel.de, martin.votava@isit.fraunhofer.de, mlan@tf.uni-kiel.de,\\ml@tf.uni-kiel.de, ars@energy.aau.dk, sssa@energy.aau.dk, fbl@energy.aau.dk
}

\maketitle

\begin{abstract}
The modular multilevel converter (MMC) is a topology that consists of a high number of capacitors, and degradation of capacitors can lead to converter malfunction, limiting the overall system lifetime. Condition monitoring methods can be applied to assess the health status of capacitors and realize predictive maintenance to improve reliability. Current research works for condition monitoring of capacitors in an MMC mainly monitor either capacitance or equivalent series resistance (ESR), while these two health indicators can shift at different speeds and lead to different end-of-life times. Hence, monitoring only one of these parameters may lead to unreliable health status evaluation. This paper proposes a data-driven method to estimate capacitance and ESR at the same time, in which particle swarm optimization (PSO) is leveraged to update the obtained estimations. Then, the results of the estimations are used to predict the sub-module voltage, which is based on a capacitor voltage equation. Furthermore, minimizing the mean square error between the predicted and actual measured voltage makes the estimations closer to the actual values. The effectiveness and feasibility of the proposed method are validated through simulations and experiments.
\end{abstract}


\section{Introduction}

Modular multilevel converter (MMC) attracts wide research attention in medium voltage and high voltage applications such as HVDC and STATCOM due to its high modularity, low harmonic component, high efficiency, etc \cite{ansariMMCBasedMTDC2020, OperationControlApplications, diabOptimalDesignControl2020, 10227452}. 
Since MMC consists of a large number of components, the reliability of critical components in the MMC should be evaluated carefully. Capacitors are one of the critical components in the MMC, as they can stabilize the submodule voltage and are key elements in voltage level generation. 
Similar to any other component,  capacitors are subjected to aging over time as they suffer from electrical and mechanical stresses during operation.
The degradation can be monitored and maintenance plan can be updated accordingly. The method of scheduling the maintenance intervention based on the health status is known as predictive maintenance, which can mitigate the loss of unplanned downtime and additional maintenance costs \cite{sundaram2022photovoltaic}. 
The key to do predictive maintenance is getting a reliable and accurate health status of the critical components, where condition monitoring methods can help.

There are different approaches to monitoring the health status of capacitors in the MMC, and the main research direction is monitoring the capacitance. This is due to the fact that the voltage and current harmonics in MMC are mainly located in the low-frequency region (below a few kHz) and the capacitance dominates the equivalent circuit model in this region, so that the other components in the equivalent circuit model, namely equivalent series resistance (ESR) and equivalent series inductance (ESL), are neglected in the capacitance calculation process \cite {polanco_condition_2022}.
The capacitance can be calculated with voltage and current harmonics, and these harmonics can come from either normal operation \cite{liu_submodule_2021,ronanki_failure_2020,ronanki_quasi-online_2019,xin_online_2020} or injected harmonics \cite{jo_condition_2014}. The capacitance can also be calculated with a capacitance-voltage relationship based on a reference submodule, where the submodule voltage ripple is inversely proportional to the capacitance when the same switching signals are given to the submodule under test and the reference submodule \cite{deng_reference_2019,yin_capacitor_2019,geng_hierarchic_2022}. It is possible that the capacitance-voltage relationship is built based on the discharging voltage curve, however, with the expenses of the exclusion of a submodule \cite{wang_condition_2019}. Using data-driven methods is another way to monitor the capacitance, e.g., the Kalman filter \cite{atkinson_new_2016} and recursive least square \cite{polanco_condition_2022}, which needs a more complicated structure than model-based methods.

Besides the capacitance, ESR in the equivalent circuit model can also be used to monitor the health status since ESR can deteriorate faster than the capacitance and reaches the end-of-life criteria much earlier than the capacitance, as reported in \cite{venetRealizationSmartElectrolytic2002, vogelsbergerLifeCycleMonitoringVoltageManaging2011,gupta_review_2018, yaobo2023Strain}. In these cases, if the condition monitoring method only monitors capacitance, the converter might operate with a degraded capacitor that has a high ESR, and the end-of-life warning is not triggered until the capacitance reaches the end-of-life criteria which makes the estimated health status unreliable. Considering the importance of monitoring ESR in reflecting the capacitor degradation, recently published research has started to monitor the ESR for capacitors in an MMC, e.g., by using the wavelet decomposition to analyze the turn-on transient voltage step which requires extra components to measure the fast and small voltage step \cite{xia_capacitor_2023} and the harmonic-based method \cite{deng_capacitor_2020}.
            
There are certain limitations in the existing methods: 1)~only monitoring a single health indicator, e.g., \cite{deng_reference_2019,yin_capacitor_2019,geng_hierarchic_2022,jo_condition_2014,liu_submodule_2021,ronanki_failure_2020,ronanki_quasi-online_2019,xin_online_2020}.
Considering the capacitance and ESR can shift at different speeds, the end-of-life time indicated by them can be different. 
For instance, if ESR degrades faster than the capacitance, the end-of-life time related to ESR will be shorter than the one related to capacitance. Between these two end-of-life times, the end-of-life warning is not triggered if only monitoring the capacitance even though the actual ESR already exceeds the end-of-life criteria.
2) the normal operation of the converter is affected during the condition monitoring due to changes in the control behaviors, e.g., \cite{deng_reference_2019,yin_capacitor_2019,geng_hierarchic_2022}.
      
To solve these problems, this paper applies a data-driven condition monitoring method to monitor capacitance and ESR together, so the health status can be derived no matter which health indicator (capacitance or ESR) degrades faster. 
The reason for using capacitance and ESR as health indicators is that these two health indicators cover the main failure mechanism, i.e., electrolyte evaporation of the aluminum electrolytic capacitor which is suitable for MMC applications \cite{wangReliabilityCapacitorsDCLink2014}. Adding more health indicators makes the condition monitoring method more complicated without covering more failure mechanisms. For instance, the weight of the capacitor decreases as the electrolyte evaporates. However, measuring the weight requires disassembling the capacitor from the print circuit board, and the weight is less sensitive than the capacitance and ESR \cite{stevens2002AECservice, yao_health_2023}.

Specifically, the proposed method uses particle swarm optimization (PSO) to update the position of particles, where the positions are estimated capacitance and ESR. Then the estimations are used to predict the submodule voltage in a sampling window based on a capacitor voltage equation. The mean square error between the predicted and measured submodule voltage is used as the objective function. The update process reduces the voltage prediction errors, and in the meantime, the estimations of health indicators are closer to the actual values. 

Compared with the existing methods, the proposed method has the advantages of: 
\begin{itemize}
    \item Monitoring both capacitance and ESR so the condition monitoring result is more reliable compared with methods based on one of the health indicators.    
    \item The method runs offline so the normal operation of the converter is not affected, and it requires no extra components. 
\end{itemize}
The system description is given in Section II; the proposed method is introduced in Section III; and validations are provided in Sections IV-V. 

\section{System Description}
This section introduces the MMC topology and operation, as well as the failure mechanism and health indicators of capacitors.
    \begin{figure}[!t]
    \centering
    \includegraphics[width=2.8 in, page=1]{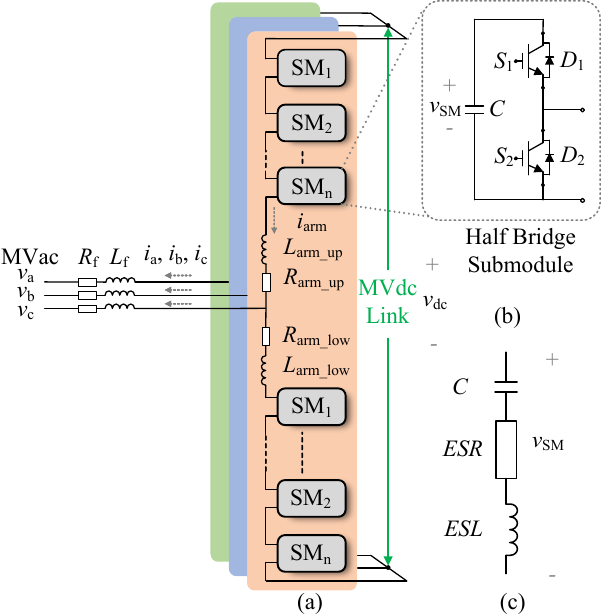}
    \caption{A modular multilevel converter, (a) Converter topology, (b) half-bridge submodule, and (c) model of the submodule capacitor with parasitic parameters.}
    \label{fig_1}
    \end{figure} 
    
\subsection{Modular Multilevel Converter (MMC)}
        Three-phase MMC with one phase in detail is depicted in Fig.~\ref{fig_1}(a). As demonstrated, three phases are connected with a medium voltage ac grid ($v_{a}$, $v_{b}$, and $v_{c}$) with a filter inductor $L_f$ and a filter resistor $R_f$. Each phase of MMC is composed of two arms, known as the upper and lower arms. A series of connections of submodules with an arm inductor $L_{arm}$ and equivalent arm resistor $R_{arm}$ forms an arm. The topology of the half-bridge submodule is shown in Fig.~\ref{fig_1}(a), which consists of a capacitor and two IGBTs with their antiparallel diode, namely $C$, $S_{1}$, $S_{2}$, $D_{1}$, and $D_{2}$, respectively.

        The MMC can be used as an inverter or an active front-end rectifier, where the outer control loop is responsible for it. In a typical MMC controller, the generated current references are considered as inputs of the inner control loop. Then, the reference of the arm voltages is produced and used in the modulation step. Apart from the mentioned controllers, MMC exploits some other controllers to improve the performance of the submodules, such as circulating current control, energy balancing control, etc. \cite{du2018modular,wu2017high}.

\subsection{Failure Mechanism of Capacitors}

The capacitor being analyzed in this section is limited to the aluminum electrolytic capacitor (AEC), which is suitable for MMC applications because of its high energy density and relatively low cost \cite{wangReliabilityCapacitorsDCLink2014}. 

The capacitor can experience electrical stress (e.g., voltage and current ripple) and thermal stress in the operation stage. For AECs, these stresses can lead to electrolyte evaporation, oxide film degradation, and anode foil degradations \cite{gupta_review_2018}.

The degradation mechanisms can be monitored with electrical or non-electrical health indicators. The electrical indicators, e.g., capacitance and ESR, can be measured online and they have a clear relation with the end-of-life criteria. While the non-electrical indicators, e.g., weight and pressure, are difficult to measure online and require a conversion stage to estimate the remaining useful lifetime. 
Therefore, using electrical indicators to estimate the health status is main research direction in capacitor condition monitoring. 

For the AEC, 
the capacitor reach the end-of-life time if one of these criteria is satisfied \cite{wangReliabilityCapacitorsDCLink2014}:
        \begin{itemize}
        \item{Capacitance reduces to $\SI{80}{\percent}$ of the initial value;}
        \item{Equivalent series resistance increases to $\SI{200}{\percent}$ of the initial value.}
        \end{itemize}

\begin{figure}[!t]
        \centering
        \includegraphics[width=3.45 in, page=2]{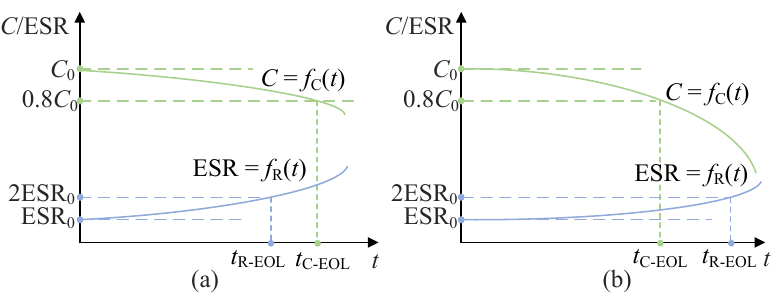}
        \caption{Health indicators of capacitors changing gradually with operation time. When the capacitance reduces to 0.8 of the initial value $C_0$ or the ESR increases to two times the original $ESR_0$, the capacitor reaches the end-of-life time $t_{C-EOL}$ or $t_{R-EOL}$, respectively. (a) ESR reaches the end-of-life criteria earlier than the capacitance, and (b) Capacitance reaches the end-of-life criteria earlier than the ESR.}
        \label{fig_2}
        \end{figure}

The equivalent circuit of the capacitor is given in Fig.~\ref{fig_1}(c), consisting of a main capacitance $C$, an equivalent series
resistance ESR, and an equivalent series inductance ESL. The capacitance and ESR affect the impedance when the frequency is below a few \SI{}{\MHz} so the ESL is negligible. The degradation process and the shifting of these two parameters are shown in Fig.~\ref{fig_2}(a)-(b), where $C$ reduces gradually from the initial value $C_0$  to \SI{80}{\percent} of $C_0$ at time $t_{C-EOL}$, and the ESR increases from the initial value $ESR_0$  to \SI{200}{\percent} $ESR_0$ at $t_{R-EOL}$.  

The figure shows also that these two indicators can vary at different speeds so they reach the end-of-life criteria at different times, i.e., $t_{C-EOL}\neq t_{R-EOL}$. 

Fig.~\ref{fig_2}(a) shows the ESR degrades faster than $C$ so $t_{R-EOL}< t_{C-EOL}$ and Fig.~\ref{fig_2}(b) shows the opposite trend so that $t_{R-EOL} > t_{C-EOL}$. The worst case is that the condition monitoring method only monitors the health indicators that degrade slower, e.g., only monitor $C$ in Fig.~\ref{fig_2}(a) so the end-of-life condition can not be detected at time $t_{R-EOL}$. From $t_{R-EOL}$ to $t_{C-EOL}$, ESR exceeds the end-of-life criteria, which can generate higher losses and lead to capacitor failure, but the end-of-life time warning can only be triggered at $t_{C-EOL}$.

\section{Condition Monitoring Method}
            \begin{figure*}[!t]
            \centering
            \includegraphics[width=6.9 in, page=15]{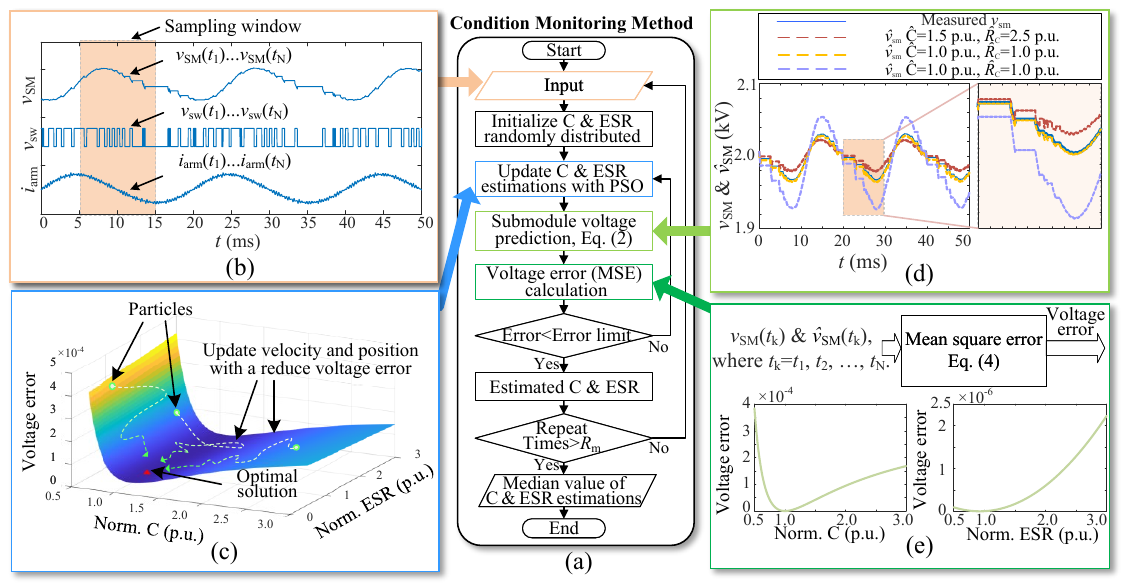}
            \caption{A system diagram of the proposed condition monitoring method:  (a) a flowchart of the condition monitoring method; (b) a sampling window of three input signals: the submodule voltage $v_{SM}(t_k)$, the switching state  $v_{sw}(t_k)$, and the arm current $i_{arm}(t_k)$, where time $t_k = t_{1}, t_{2},..., t_{N}$ consists of $N$ points; (c) the particle swarm optimization (PSO) method updates the positions of each particle to have a low voltage error; (d) the voltage prediction module predicts the submodule voltage $\hat{v}_{SM}(t_k)$ in the sampling window, where the prediction with 1 p.u. estimations is closer to the measured voltage; (e) the mean square error between predicted and measured voltage error is calculated, and it is minimized when both C and ESR are 1 p.u. 
            }
            \label{fig:CM_flowchart}
            \end{figure*}

            To overcome the limitations in the previous capacitor condition monitoring methods, this paper proposes a method to estimate the two health indicators (being capacitance and ESR) together, while not affecting the converter operation. 
            
            The system diagram of the proposed method is illustrated in Fig. \ref{fig:CM_flowchart}(a).          
            The proposed method starts with taking input signals and initializing capacitance and ESR which are randomly distributed in the solution space. Then the capacitance and ESR are updated with particle swarm optimization, and the updated estimations are used to predict the submodule voltage within a sampling window based on the capacitor voltage equation. The mean square error between the predicted and measured voltage is calculated and compared with an error limit to determine if the update should continue. The condition monitoring method is repeated multiple times ($R_m$) and the median values of these estimations are taken as the final estimations to avoid using an outlier as the final estimations.
            The key modules in the flowchart are introduced in the following sections.

        \subsection{Input Signal}
            Three signals related to one submodule being monitored are measured, including the submodule voltage $v_{SM}(t_k)$, the switching state of the submodule $v_{sw}(t_k)$, and the arm current $i_{arm}(t_k)$.   
                        
            An example waveform is shown in Fig. \ref{fig:CM_flowchart}(b). These three signals are time series in a sampling window, $t_k = t_{1}, t_{2},..., t_{N}$.
            The sampling window is selected as half of the fundamental cycle within which there are some switching transients because it is important for ESR estimation \cite{xia_capacitor_2023}, while the window is shorter than a fundamental cycle to mitigate the error in the voltage prediction process. The sampling frequency is \SI{100}{\kHz}, which is much higher than the switching frequency of an MMC (usually lower than several \SI{}{\kHz} \cite{du2014modulation}) to ensure the prediction accuracy.            
            
        \subsection{Update C and ESR with Particle Swarm Optimization}
            Particle swarm optimization (PSO) is a classical metaheuristic search method. It searches the solution space with a group of particles and the positions of the particles are updated gradually with shared information to find the optimal solution. 

            In the proposed method, each particle has a 2-dimensional position: estimated capacitance and ESR. These two values are assigned randomly within the solution space during the initialization stage, marked as green circles in Fig. \ref{fig:CM_flowchart}(c). Each particle also has a randomly given initial velocity. The position of each particle leads to an individual voltage error.    
            In each iteration, particles update their velocities and positions \cite{clercParticleSwarmOptimization2006}:  
            \begin{equation}
            \label{eq:psov}            
            \begin{split}            
              v_{j} = \omega_j v_{j-1} + c_{1}r_{1j}(x_{p}-&x_{j-1}) + c_{2}r_{2j}(x_{g}-x_{j-1}) \\
              x_{j} = &x_{j-1} + v_{j}          
            \end{split}
            \end{equation}
            where $v_{j}$ and $v_{j-1}$ are velocities, $x_{j}$ and $x_{j-1}$ are positions of the $j^{th}$ and $(j-1)^{th}$ iteration, respectively; $x_{p}$ and $x_{g}$ are individual and global optimal positions; $\omega_j$, $c_1$, and $c_2$ are inertia weight, cognitive weight and social weight, respectively; $r_{1j}$ and $r_{2j}$ are two random numbers between 0 and 1, varying in each iteration. The inertia weight $\omega_j$ decreases gradually so that as the iteration increases, the velocity reduces and the searching step becomes more refined \cite{clercParticleSwarmOptimization2006}. The reducing velocity can be seen on the trajectory, which moves in a wider range at the beginning and becomes more stable at the end. Three simplified trajectories are plotted in Fig. \ref{fig:CM_flowchart}(c) as green dashed lines.

            Eq. \ref{eq:psov} shows that the positions are updated according to the global and individual optimal solution, so the voltage errors of all particles reduce gradually and the positions of particles move to the optimal solution gradually. In other words, the update process uses particles to approach the actual values.
            
        \subsection{Voltage Prediction and Error Calculation}   
            The voltage prediction block takes the measured signals (i.e., $i_{arm}(t_k)$ and $v_{sw}(t_k)$) as well as the estimated parameters (i.e., $\hat C$ and $\hat R_{C}$) to predict a series of submodule voltage $\hat v_{SM}(t_k)$, where $t_k=t_1,t_2,...,t_N$ covering the entire sampling window. 
            At the first time step $t_1$, $\hat v_{SM}(t_{1})  = v_{SM}(t_{1})$. The prediction starts from the second time step $t_2$ to the last time step $t_N$ in the sampling window, and the prediction is based on a capacitor voltage equation:
                   
            \begin{equation}
            \label{eq:vsm}
            \begin{split}
             \hat v_{SM}(t_{k})  = &  \hat v_{SM}(t_{k-1}) + \frac{i_{c}(t_{k})T_s}{\hat C}+ {\hat R_{C}}[i_{c}(t_{k})-i_{c}(t_{k-1})]   
            \end{split}
            \end{equation}        
            where $T_{s}$ is the sampling time; $i_{c}(t_k)$ is the capacitor current and it is the product of the arm current and the switching state:
            \begin{equation}
            \label{eq:vsm2}
            \begin{split}
             i_{c}(t_k)  = i_{arm}(t_k)v_{sw}(t_k)
            \end{split}
            \end{equation}          
            where the switching state $v_{sw}(t_k)$ is one when the upper switch of the 
            submodule is turned on and the capacitor is in the charging/discharging state; while $v_{sw}(t_k)$ is zero when the lower switch in the submodule is turned on and the capacitor is bypassed.
        
            The prediction is evaluated with the mean square error between the predicted submodule voltage $\hat v_{SM}(t_{k})$ and the measured voltage $v_{SM}(t_{k})$. 
            \begin{equation}
            \label{eq:cost}
             V_{err} = \frac {1}{{V_{m}^2}N}\sum _{t_k=t_1}^{t_N}(v_{SM}(t_k) - \hat {v}_{SM}(t)_k)^{2}
            \end{equation}            
            where ${V_{m}}$ is the maximum measured submodule voltage within the sampling window and $N$ is the number of time steps in the sampling window.
         
            The submodule voltage error is compared with an error limit to determine if the capacitance and ESR estimations should be updated. A large error indicates that the estimations deviate significantly from the actual value while a small error reveals that the estimations are close to the actual values. 
            Fig.\ref{fig:CM_flowchart}(d) shows the voltage prediction waveforms. When the estimated capacitance and ESR are different from 1 p.u., the prediction waveforms in dashed lines significantly deviate from the measured voltage (a solid blue line). The voltage error against the normalized capacitance and ESR when their counterparts are 1 p.u. are given in Fig.~\ref{fig:CM_flowchart}(e), which shows that the voltage error is minimized when both estimations achieve 1~p.u. 

\section{Simulation Study}
This section validates the design, which is the selection of swarm size and error limit for the data-driven condition monitoring method, and the result which includes voltage predictions and health indicator estimations.
\subsection{Simulation Setup}
\begin{figure}[!t]
            \centering
            \includegraphics[width=3.3 in, page=4]{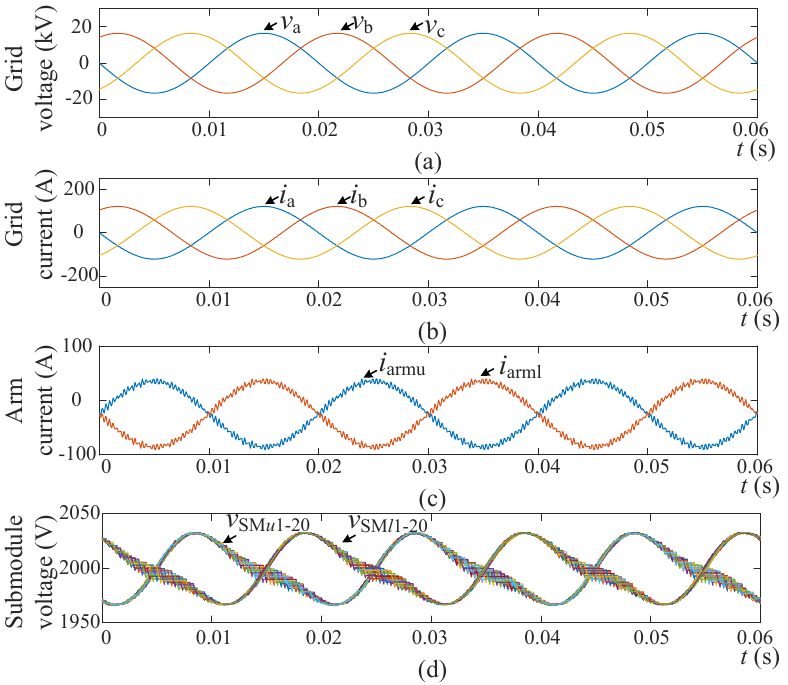}
            \caption{MMC simulation waveforms of (a)three-phase grid voltage, (b) three-phase arm current, (c) arm current in phase A, (d) submodule voltage of the upper and lower arm in the phase a.}
            \label{fig_simwave1}
            \end{figure}

    The condition monitoring method is validated with a three-phase MMC simulation model built in PLECS. The circuit model and parameters are listed in Table \ref{tab:lab_MPE} and the MMC converter is operating in the rectifier mode at full load, i.e., $P_{in} = \SI{3}{\MW}$. The key waveforms are illustrated in Fig. \ref{fig_simwave1}. The current is regulated by a current controller where it can be seen that the grid current ($i_a$, $i_b$, and $i_c$) is in phase with the phase voltage ($v_a$, $v_b$, and $v_c$). The upper and lower arm currents ($i_{armu}$ and $i_{arml}$) are balanced by a circulating current controller. The submodule voltage is controlled with a sorting method to balance the submodule voltages of 20 submodules in one arm ($v_{SMu1-20}$ and $v_{SMl1-20}$). 

    After running the simulation model, the signals of the submodule voltages, arm current, and switching signals are recorded for condition monitoring purposes. 

        \begin{table}[!t]
        \scriptsize
        \caption{Simulation and experimental setup parameters.}\label{tab:lab_MPE}
        \centering
        \begin{tabular}{|c|c|c|c|}
        \hline
        \textbf{Parameter}                     & \textbf{Var.} & \textbf{Simulation} & \textbf{Experimental}      \\ \hline
        Input power                   & $P_{in}$      & \SI{3}{\MW}     &  N/A         \\ \hline
        DC voltage                    & $V_{DC}$     & \SI{40}{\kV}     & \SI{200}{\V}        \\ \hline
        AC voltage                    & $V_{AC}$       & \SI{16.5}{\kV}    & N/A          \\ \hline
        AC frequency                  & $f_{g}$        & \SI{50}{\Hz}      & \SI{50}{\Hz}        \\ \hline
        SM switching frequency        & $f_{sw}$      & \SI{3}{\kHz}      & \SI{1}{\kHz}        \\ \hline
        Full bridge switching frequency        & $f_{cg}$      & N/A      & \SI{10}{\kHz}        \\ \hline
        Filter inductance             & $L_{f}$       & \SI{40}{\mH}      &  \SI{5.4}{\mH}             \\ \hline
        Arm inductance                & $L_{arm}$     & \SI{10}{\mH}     &   N/A           \\ \hline
        Filter resistance             & $R_{f}$       & \SI{100}{\mohm}      &  \SI{1}{\ohm}            \\ \hline
        Arm resistance                & $R_{arm}$     & \SI{100}{\mohm}      &   N/A           \\ \hline
        Number of SM per arm          &  -        & 20         & 1            \\ \hline
        SM capacitance                & $C$         & \SI{2.2}{\mF}     & \SI{2.26}{\mF}     \\ \hline
        SM capacitance ESR            & ESR         & \SI{40}{\mohm}    & \SI{44.12}{\mohm}   \\ \hline
        Sampling window            & -        &  \SI{10}{\ms}    &  \SI{10}{\ms}     \\ \hline
        Sampling frequency            & $f_{sa}$        &  \SI{100}{\kHz}    &  \SI{100}{\kHz}     \\ \hline
        Voltage loop bandwidth        & $f_{v}$        & N/A    &  \SI{20}{\Hz}  \\ \hline
        Current loop bandwidth        & $f_{i}$        & \SI{800}{\Hz}     & \SI{800}{\Hz}      \\ \hline
        Repeat times        & $R_{m}$      & 15        & 15  \\ \hline
        Swarm size              & -        & 10    & 10   \\ \hline            
        Error limit              & -        & $10^{-6}$    & $10^{-6}$   \\ \hline
        Maximum iteration         & -        & $100$    & $100$   \\ \hline
        Cognitive weight                     & $c_1$       & 1.49    & 1.49      \\ \hline
        Social weight                     & $c_2$        & 1.49    & 1.49      \\ \hline
        Boundary of $C$ estimation        & -        & $[1.1, 6.6]$\SI{}{\mF}    & $[1.1, 6.6]$\SI{}{\mF}   \\
        \hline
        Boundary of $ESR$ estimation        & -        & $[20, 120]$\SI{}{\mohm}    & $[20, 120]$\SI{}{\mohm}    \\
        \hline
        \end{tabular}
        \end{table}

\subsection{Parameter Design of Data-driven Condition Monitoring}
            \begin{figure}[!t]
            \centering
            \includegraphics[width=3.05 in, page=11]{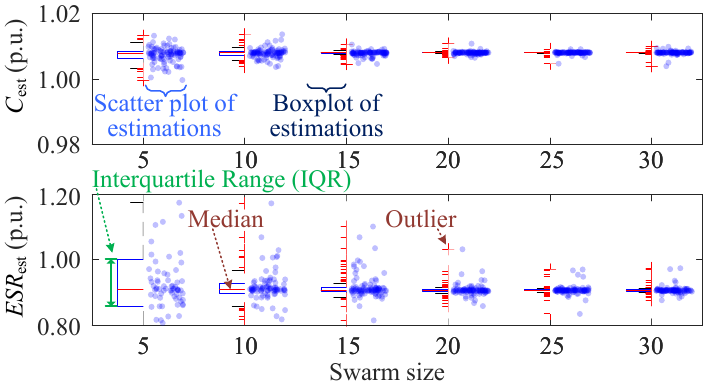}
            \caption{The effect of the swarm size on estimation results distribution.}
            \label{fig_swarmsize}
            \end{figure}            
            \begin{figure}[!t]
            \centering
            \includegraphics[width=3.05 in, page=12]{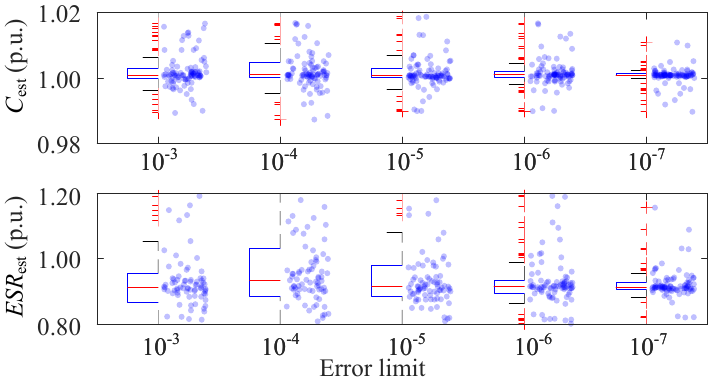}
            \caption{The effect of the error limit on estimation results distribution.}
            \label{fig_tolerance}
            \end{figure}            
            To tailor the optimization method for condition monitoring, the effect of two parameters is studied: the number of swarms (also named the swarm size), and the error limit.

            The swarm size is usually between 20 and 30 \cite{clercParticleSwarmOptimization2006} therefore 30 is the maximum swarm size to be studied. The test conditions are listed in Table \ref{tab:lab_MPE}, and the capacitor has capacitance and ESR at their initial values, 1~p.u. The condition monitoring method repeats for a hundred times to visualize the distribution of estimations, as shown in Fig. \ref{fig_swarmsize}. 

            The scatter plot and the boxplot illustrate the distribution of estimations. With an increasing swarm size, the estimations (blue dots) concentrate gradually. The box plots also illustrate the concentration of estimations; the height of the box (named the interquartile range or IQR) covers most of the estimations and reduces gradually with increased swarm sizes. For instance, if the swarm size is 5, the interquartile ranges of capacitance and ESR estimations are $\SI{0.21}{\percent}$ and $\SI{14.47}{\percent}$ of 1~p.u., respectively; while if the swarm size is 10, these values are $\SI{0.11}{\percent}$ and $\SI{2.95}{\percent}$, respectively. Considering the acceptable error ranges for $C$ and ESR are $\pm\SI{1}{\percent}$ and $\pm\SI{10}{\percent}$, respectively \cite{deng_capacitor_2020}, the IQR of ESR when the swarm size is 5 is $\SI{14.47}{\percent}$ which is higher than the acceptable error range $\pm\SI{10}{\percent}$. Therefore, it is better to have a larger swarm size, and the swarm size is selected as ten, which gives stable estimations in the acceptable error range.

            The second parameter that needs to be considered is the error limit. The error limit is usually in the range of $10^{-3}$ and $10^{-7}$, and its effect on estimations and iterations are shown in Fig. \ref{fig_tolerance}. When the error limit reaches $10^{-6}$, both capacitance and ESR estimations concentrate around the median value, with interquartile ranges of $\SI{0.19}{\percent}$ and $\SI{3.86}{\percent}$, respectively. Therefore, the error limit is selected as $10^{-6}$.     
            
\subsection{Voltage Prediction}
            The voltage prediction function is validated in the full load condition with two capacitors: a healthy capacitor ($C$ = ESR~ = 1 p.u.) and a fully degraded capacitor ($C$ = 0.8 p.u. and ESR~= 2.0 p.u.).

\begin{figure}[!t]
            \centering
            \includegraphics[width=3.3 in, page=5]{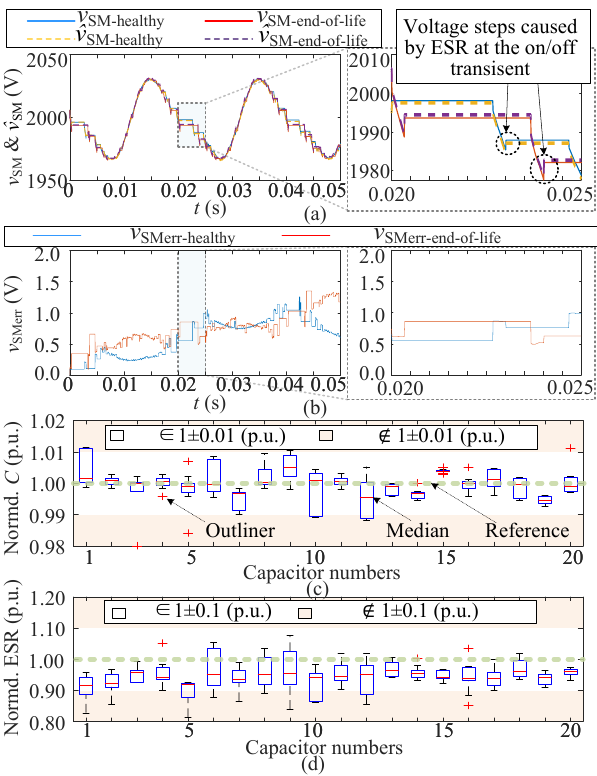}
            \caption{Simulation results: (a) waveforms of the sensed submodule voltage $v_{SM}$ (solid lines) and predicted submodule voltage $\hat v_{SM}$ with different capacitances and ESRs (dashed lines); (b) the instantaneous error between $v_{SM}$ and $\hat v_{SM}$; (c) and (d) are boxplots of the estimated capacitances and ESRs.
            }
            \label{fig_simresult1}
            \end{figure}

            The sensed voltage is compared with the predicted voltage in Fig. \ref{fig_simresult1}(a). The voltage prediction in dashed lines can follow the voltage measurement in solid lines with an instantaneous voltage error below \SI{2}{\V}, which is $\SI{0.1}{\percent}$ of the $\SI{2}{kV}$ average submodule voltage, as seen in Fig. \ref{fig_simresult1}(b). In the zoomed-in waveform on the right side, voltage steps caused by current steps during the switching transient can be seen. The magnitude of the voltage step is proportional to the ESR (Eq. \eqref{eq:vsm}) so the degraded capacitor has a higher voltage step because of the higher ESR.

            Considering the voltage error is below \SI{2}{\V} or \SI{0.1}{\percent} of the \SI{2}{\kV}, the estimation error is acceptable and should only affect the capacitance and ESR estimations in a limited range. 
            
\subsection{C and ESR Estimation}
    This section validates the function of capacitance and ESR estimation with different health statuses. The test condition is that $C$ and ESR vary linearly between the healthy condition and the end-of-life criteria: 
            \begin{itemize}
                \item $C$: 1, 0.99, ..., 0.8 p.u.
                \item ESR: 1, 1.05, ..., 2 p.u.
            \end{itemize}
    The solution space is half and three times the nominal value (i.e., $\hat{C}$ $\in$ [0.5,3] p.u. and $\hat{R_C}$ $\in$ [0.5,3] p.u.) to cover both the end-of-life criteria and a $\pm\SI{20}{\percent}$  tolerance.           
            
     The condition monitoring method repeats 15 times for 20 submodules in the upper arm of phase A, and all normalized estimations are illustrated with boxplots in Figs. \ref{fig_simresult1}(c) and (d). The acceptable estimation ranges, $1\pm\SI{1}{\percent}$ and $1\pm\SI{10}{\percent}$ for capacitances and ESRs, respectively, are colored with a white background, while the other ranges are marked with a light red background. The median values of estimations are red lines, and a green line at 1 p.u. is added as a reference. The median values for the 20 monitored capacitors are in the acceptable range.  In total, the average estimation errors of capacitance and ESR are $\SI{0.18}{\percent}$ and $\SI{5.47}{\percent}$. 

    The boxplot can also show the distribution of all estimations; the blue box represents the $\pm\SI{25}{\percent}$ range around the median value. For the capacitance and the ESR estimations, most of the blue boxes are in the ranges of $\pm\SI{1}{\percent}$ and $\pm\SI{10}{\percent}$, which shows that the condition monitoring method can have acceptable estimations during most of the iterations. 
       
\section{Experimental Result}
\subsection{Test Setup}
    Experiments are carried out with a mission profile emulator (MPE). The MPE is a simplified reliability test setup; it emulates a mission profile for the submodule under test similar to the full MMC operation. It can reduce the required testing time and cost for building and testing an entire MMC converter because an MMC usually has a large number of components and complicated control schemes.            

\begin{figure}[!t]
            \centering
            \includegraphics[width=2.65 in, page=6]{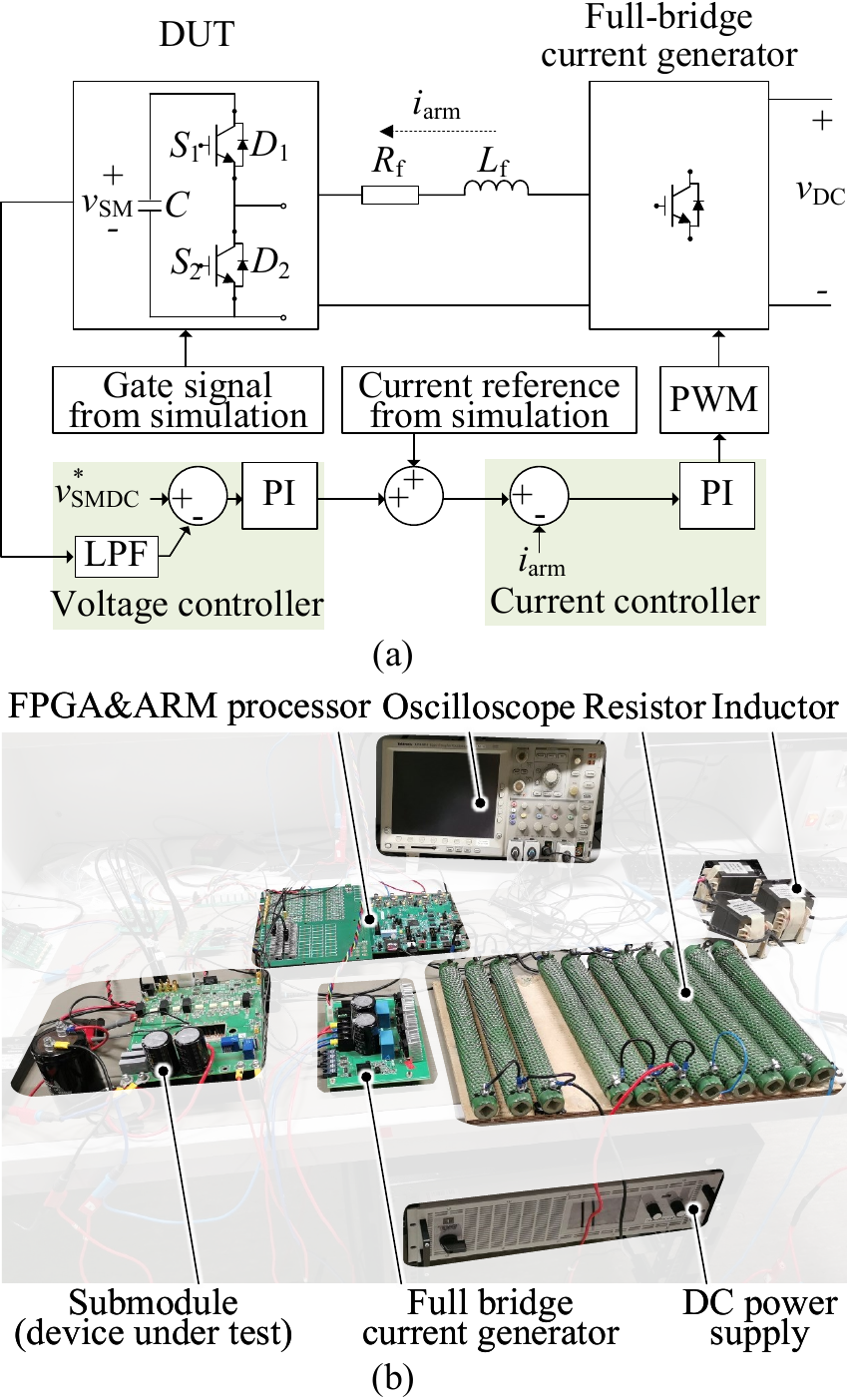}
            \caption{Mission profile emulator test setup: (a) Topology and control diagram of the mission profile emulator used to validate the method. (b) Test setup.}
            \label{fig_test_setup}
            \end{figure}

    The mission profile emulator MPE parameters are listed in the experimental column of Table \ref{tab:lab_MPE} and the control diagram is illustrated in Fig.~\ref{fig_test_setup}(a). The MPE includes a DC voltage supply, a full-bridge current generator, a filter inductor $L_f$, a filter resistor $R_f$, and a submodule for testing (also named the device under test or DUT). There are two control loops in the control diagram, a current control loop and a voltage control loop. The current controller gets the reference current 
    from the simulation and regulates the filter current $i_{arm}$ by controlling the full-bridge current generator; while the voltage controller maintains the DC component in the submodule voltage $v_{SM}$ by adding an offset $v_{SMDC}^*$ to the voltage reference. Therefore, the filter current $i_{arm}$ includes a DC component and an AC component.
    
    The MPE test setup is illustrated in Fig. \ref{fig_test_setup}(b). The submodule voltage $v_{SM}$ and the filter current $i_f$ are sensed with a voltage divider and a current sensor LEM CKSR 15NP, respectively. These signals are sent to the controller implemented on an FPGA and an ARM processor on a ZedBoard for control purposes. They are also sent to the oscilloscope and saved for condition monitoring validation. The DUT is connected to a group of three aluminum electrolytic capacitors. The overall capacitance is \SI{2.26}{\mF} and the ESR is \SI{44.12}{\mohm} which are measured with an impedance analyzer Keysight E4990A. 

\subsection{Experimental Result and Discussion}
            \begin{figure}[!t]
            \centering
            \includegraphics[width=2.65 in, page=9]{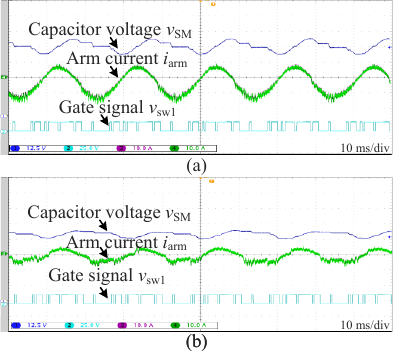}
            \caption{Experimental waveforms: (a) test scenario 1 and (b) test scenario 3.}
            \label{fig_lab_waveform1}
            \end{figure}

            \begin{table}[!t]
            \caption{Experimental test conditions and estimation errors of capacitance and ESR.}\label{tab:lab_err}
            \centering
            \begin{tabular}{|c|c|c|c|c|c|c|}      
            \hline
            \textbf{Test scenario} & \textbf{1} &\textbf{2} & \textbf{3} & \textbf{4} & \textbf{5} & \textbf{6}\\
            \hline
            $V_{SMDC}^*$ (V) & \multicolumn{3}{c|}{50 V}  & \multicolumn{3}{c|}{30 V} \\
            \hline
            ${I}_{AC}^*$ (A)  & 9 & 6 & 3 & 9 & 6 & 3\\
            \hline
            ${C}_{err}$ ($\%$)  & -0.65 & -0.63 & 1.84 & -0.43 & 1.02 & 2.39\\
             \hline
            ${ESR}_{err}$ ($\%$)  & 2.28 & -4.46 & -4.45 & 5.01 & 8.15 & 5.85\\           
            \hline
            \end{tabular}
            \end{table}   

            \begin{figure}[!t]
            \centering
            \includegraphics[width=3.1 in, page=10]{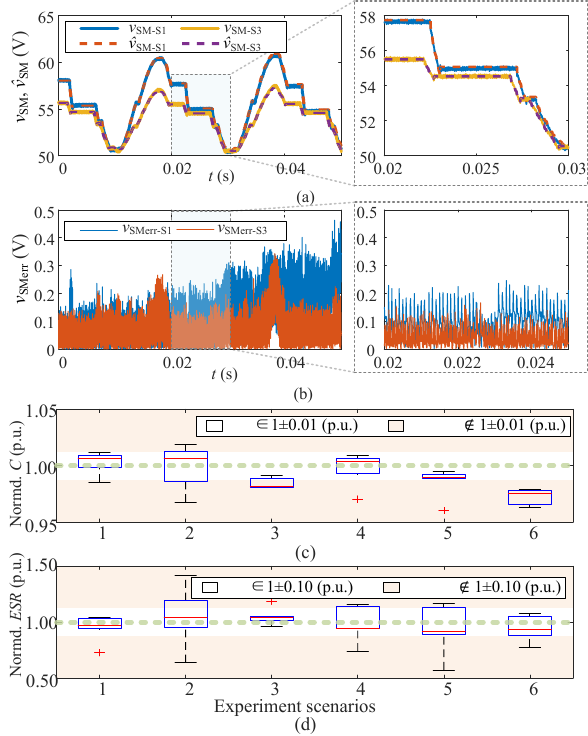}
            \caption{Comparison between the measured submodule voltage and the predicted submodule voltage. (a) The measured and estimated submodule voltage waveforms for scenario 1 and scenario 3 in $\SI{0.05}{\s}$; (b) The voltage error between the measurement and estimation; (c) and (d) are boxplots of capacitance and ESR estimations in six scenarios. respectively.}
            \label{fig_lab_waveform3_predict}
            \end{figure}

             The experiments are designed to validate that the proposed method can provide accurate estimations under different load conditions. The testing scenarios are listed in Table \ref{tab:lab_err}, showing the DC component of the submodule voltage varies from \SI{50}{\V} to \SI{30}{\V}, and the magnitude of the AC reference current varies between \SI{9}{\A}, \SI{6}{\A}, and \SI{3}{\A}. The operating waveforms for the first and the third test scenarios, including the submodule voltage $v_{SM}$, the filter current $i_{f}$, and the switching signal of the upper switch $v_{sw1}$ are shown in Fig. \ref{fig_lab_waveform1}.
            
            To verify the function of the voltage prediction block, the predicted submodule voltage $\hat v_{SM}$ is compared with the voltage measurement $v_{SM}$, as shown in Fig. \ref{fig_lab_waveform3_predict}. The voltage error $v_{SMerr}$ is the instantaneous difference between the prediction $\hat v_{SM}$ and the measurement $v_{SM}$. As seen in Fig.~\ref{fig_lab_waveform3_predict}(c), $v_{SMerr}$ increases as the prediction time increases, but the error is lower than \SI{0.5}{\V} within a $\SI{0.05}{\s}$ period.

            The normalized capacitance $C$ and ESR estimations are given in Figs.~\ref{fig_lab_waveform1}(c) and (d), wherein the green dashed line is the reference value; the red line is the median estimation; the blue box represents the $\pm\SI{25}{\percent}$ estimations; the whisker above and under the box means outliers of estimations; the red areas are the range above $\pm\SI{1}{\percent}$ and $\pm\SI{10}{\percent}$ for the $C$ and ESR, respectively. Except for test scenario 3 and test scenario 6, the median values of estimated capacitance and ESR are around $\pm\SI{1}{\percent}$ and $\pm\SI{10}{\percent}$, respectively. The higher error of $C$ for test scenarios 3 and 6 can come from the small voltage ripple as shown in Fig. \ref{fig_lab_waveform1} (b).  These estimation errors are listed in Table \ref{tab:lab_err}. Compared with simulation results in Figs. \ref{fig_simresult1}(c) and (d), the experimental results in Fig. \ref{fig_lab_waveform1} have a wider blue box, meaning that the results have higher oscillations.

\section{Conclusion}
Based on the fact that health indicators can change at different speeds and lead to different end-of-life times, this paper proposes a condition monitoring method that can monitor both capacitance and ESR, which makes the condition monitoring results more reliable compared with other methods that monitor only one of them. The method combines the particle swarm optimization and the capacitor voltage prediction functions. The design of key parameters (swarm size and error limit) is studied to guarantee that the estimation errors of capacitance and ESR are in the expected ranges. The effectiveness of the method is validated in the simulations and experimental stages. Except for light-load cases, the estimation errors of the capacitance and ESRs are around  $\SI{1}{\percent}$ and $\SI{10}{\percent}$, respectively.

\section*{Acknowledgment}
This project is supported by the European Union’s Horizon 2020 research and innovation program under the Marie Skłodowska-Curie grant agreement No. 955614.

\bibliographystyle{ieeetr}
\bibliography{c2ref}

\vspace{12pt}

\end{document}